\documentclass[aps,prl,twocolumn,nofootinbib,amsfonts,pdftex,floatfix,superscriptaddress]{revtex4-1}

\usepackage{graphicx}  
\usepackage{amsmath, amssymb, bm}   

\usepackage[dvips]{color}
\usepackage{natbib}
\usepackage{xcolor}
\usepackage{hyperref}
\hypersetup{
    colorlinks,
    linkcolor={red!50!black},
    citecolor={blue!50!black},
    urlcolor={blue!80!black}
}

\usepackage{slashed}

\def\){\right)} 
\def\({\left(} 
\def\]{\right]} 
\def\[{\left[}
\def\nopi{$\slashed{\pi}$EFT}
\def\Trel{\bm{T}_\mathrm{rel}}

\begin{document}

\title{Universal behavior of $p$-wave proton-proton fusion near threshold}

\author{%
Bijaya Acharya}
\email{acharya@uni-mainz.de}
\affiliation{Institut f\"{u}r Kernphysik and PRISMA$^+$ Cluster of Excellence, Johannes Gutenberg-Universit\"{a}t Mainz, 55128 Mainz, Germany}
\affiliation{Department of Physics and Astronomy, University of
  Tennessee, Knoxville, TN 37996, USA}

\author{%
Lucas Platter}
\email{lplatter@utk.edu}
\affiliation{Department of Physics and Astronomy, University of
  Tennessee, Knoxville, TN 37996, USA}
\affiliation{Physics Division, Oak Ridge National Laboratory, Oak
  Ridge, TN 37831, USA}

\author{%
Gautam Rupak}
\email{grupak@ccs.msstate.edu}
\affiliation{Department of Physics \& Astronomy and HPC\textsuperscript{2} 
Center for Computational Sciences,
Mississippi State University, Mississippi State, MS 39762, USA}

\begin{abstract}
  We calculate the $p$-wave contribution to the proton-proton fusion
  $S$ factor and its energy derivative in pionless effective field
  theory (EFT) up to next-to-leading order. The leading contributions are
  given by a recoil piece from the Gamow-Teller and Fermi operators, 
  and from relativistic $1/m$ suppressed weak interaction operators.
  We obtain the value of $(2.5\pm0.3 )\times 10^{-28}~\mathrm{MeV\ fm^2}$ for
  the $S$ factor and $(2.2\pm0.2) \times 10^{-26}~\mathrm{fm^2}$ for its energy
  derivative at threshold. These are smaller than the results of a
  prior study that employed chiral EFT by several
  orders of magnitude. We conclude that, contrary to what has been
  previously reported, the $p$-wave contribution does not need to be considered in a high-precision determination of the $S$ factor at
  astrophysical energies. Combined with the chiral EFT calculation of Acharya {\it et al.} [\href{https://doi.org/10.1016/j.physletb.2016.07.032} {Phys. Lett. B {\bf 760}, 584 (2016)}] for the $s$-wave channel, this gives a total threshold $S$ factor of $S(0) = (4.047^{+0.024}_{-0.032}) \times 10^{-23}~{\rm MeV~fm}^2$.
  
\end{abstract}
 
\maketitle


The Sun is powered by nuclear burning of hydrogen, the most abundant
element in the universe, into helium. The elementary proton-proton
$(pp)$ fusion process that results in a deuteron, a positron and a
neutrino is the first step in the chain of reactions producing heavier
elements in stellar environments~\cite{Bethe:1938yy}. Solar models for
quantities such as core temperatures and neutrino flux are sensitive
to the $pp$ fusion cross section.  At the relevant solar core
temperatures ($T\approx 1.5\times10^7$ K), the Coulomb repulsion and the
slow weak process result in a very small cross section. Thus,
experimental measurements are prohibitive and
non-existent. Theoretical calculations with well-justified uncertainty
estimates are essential for providing critical input data for stellar
models~\cite{Adelberger:1998qm,Adelberger:2010qa,Vinyoles:2016djt}.
Inference of solar neutrino masses from terrestrial measurements
depends crucially on the $pp$ fusion rate. This reaction involves all
the fundamental interactions except gravity. It is important in the
field of astro, nuclear and particle physics, and there is an active
effort to calculate the cross section with ever higher accuracy and
precision [see Reference~\cite{Adelberger:2010qa} for an extensive
review of the existing literature].

The reaction cross section $\sigma(E)$ at center-of-mass (c.m.) kinetic energy $E$ is
conventionally expressed in terms 
of the $S$ factor $S(E)=E \exp(2 \pi \eta_p)\sigma(E)$.  The Sommerfeld
parameter $\eta_p=\sqrt{m_p/E}\, \alpha/2$ with proton mass
$m_p=938.28$ MeV and fine structure constant $\alpha=1/137$.  Reference~\cite{Adelberger:2010qa}
provides the best estimates of
$S(0)=(4.01 \pm 0.04)\times 10^{-23}~{\rm MeV~fm}^2$ at threshold, and
$S^\prime(0)/S(0)=(11.2 \pm 0.1)~{\rm MeV}^{-1}$ for the logarithmic
derivative. Reference~\cite{Adelberger:2010qa} also estimated the contribution of the
$S^{\prime\prime}(0)$ term to be $\approx1\%$ at the solar core
temperature and recommended that a modern calculation be
undertaken. The threshold $S$ factor and its energy derivatives have
since been calculated in pionless~\cite{Chen:2012hm} and
chiral~\cite{Marcucci:2013tda,Acharya:2016kfl} effective field
theories (EFTs). Reference~\cite{Marcucci:2013tda}, the only study so
far to have included capture from the $p$-wave, has claimed that this
channel makes a significant contribution to $S(E)$, of roughly the
same size as the $s$-wave $S^{\prime\prime}(0)$ term, in the
astrophysically relevant $E\approx10$~keV region.\footnote{After the authors of Reference~\cite{Marcucci:2013tda} were notified about this Rapid Communication, they revisited their calculation and published an Erratum~\cite{Phys.Rev.Lett.123.019901} whose $p$-wave result, albeit much closer,  is still not in agreement with this work within our uncertainty estimate. More importantly, as we will later discuss, the result of Reference~\cite{Phys.Rev.Lett.123.019901} for the total $S$ factor, with $s$ and $p$ waves included, does not agree with the value we quote here due to basis-truncation issues in their calculation.} An independent
calculation of the $p$-wave contribution is therefore imperative,
especially since the $s$-wave $S(E)$ has now been constrained to
subpercentage precision~\cite{Acharya:2016kfl}.

EFTs provide a description of interacting particles in terms of only
those degrees of freedom that are relevant below a breakdown momentum
scale, $\Lambda$. Low-energy observables are then calculated as expansions in powers of $Q/\Lambda$, where $Q$ is the characteristic
momentum of the process under study. Such approaches have been widely
used in nuclear physics.  They provide a clear guidance on how to
systematically construct the nuclear Hamiltonian and couplings to
external electroweak sources as perturbation in $Q/\Lambda$.  They
also enable us to use the convergence of the expansion to estimate the
uncertainty in theoretical calculations.  The $pp$ fusion process at
solar energies $E\lesssim 100$ keV is peripheral, and thus can be
accurately described in terms of the incoming $pp$ $s$-wave phase shift
and the outgoing deuteron bound state wave function to within 10\%
model-independently~\cite{Bethe:1938yy}. Thus the characteristic
momentum scale $Q \approx p, \gamma, 1/a_{pp}, \alpha m_p \ll m_\pi$,
where $p=\sqrt{m_pE}\lesssim 10$ MeV, $a_{pp}\approx 25$ MeV is the $pp$
$s$-wave scattering length, $\gamma=45.701$ MeV the deuteron binding
momentum, $m_\pi\approx 140$ MeV the pion mass. It is therefore
appropriate to employ Pionless EFT (\nopi) for the calculation of the
$pp$ fusion $S$ factor. This is an EFT with non-relativistic nucleons
that interact through short-ranged forces without an explicit pion
degree of freedom~\cite{vanKolck:1998bw,Chen_1999}. Its breakdown
scale is $\Lambda\approx m_\pi$ and the perturbative expansion is
therefore in $Q/\Lambda\lesssim1/3$. \nopi~provides a simple
description of $pp$ fusion, to about 10\% precision, in terms of
nucleon-nucleon observables~\cite{Kong:1999tw}. Calculation of the
fusion rate to a few percent precision requires contribution from
two-body currents that represent short-distance physics not
constrained by elastic-channel nucleon-nucleon phase
shifts~\cite{Butler:2001jj}.

In this Rapid Communication, we present the first calculation of the $p$-wave
contributions to $pp$ fusion in \nopi.  The results are expressed in
terms of model-independent parameters, and, therefore, universal. It
provides an important constraint on the precise determination of the
solar $pp$ fusion rate and provides insights into further steps that
are needed to reduce uncertainties in the future.

\paragraph{\bf Pionless effective field theory:}
\label{sec:pionl-effect-field}
The cross section calculation depends on the strong interaction, the
Coulomb repulsion between the two protons, and the weak
interaction. The dominant $p$-wave contribution requires the strong
interaction only in the outgoing deuteron ($^{3}S_1$) channel, which
is given by~\cite{vanKolck:1998bw,Kaplan:1998we, Kaplan:1998tg, KSW99}
\begin{align}
  \label{eq:Lagrangian-pionless}
  \mathcal{L}_S =  \,
   & d_i^\dagger\left[ \Delta -\left(i
      \partial_0+\frac{\nabla^2}{4m}\right)\right]d_i  
  \nonumber\\
   & + g_0\left[ d_i^\dagger(N^TP_i N) +\rm{h.c}\right]\, ,
\end{align}
where $m=938.92$ MeV is the isospin-averaged nucleon mass, $N$
represents a nucleon and the vector $d_i$ represents the
deuteron. $P_i=\sigma_2\sigma_i\tau_2/\sqrt{8}$, where the Pauli
matrices $\bm{\sigma}$ and $\bm{\tau}$ respectively act on spins and
isospins, projects the nucleons onto the spin-triplet isospin-singlet
$^{3}S_1$ (deuteron) channel. The two couplings $\Delta$, $g_0$ are
fixed by requiring that the deuteron bound state wave function has the
correct exponential decay and normalization constant. In \nopi, this
corresponds to ensuring the $^{3}S_1$ elastic scattering amplitude has
a pole at $p^\ast=i\gamma$, and has the correct residue at the said
pole.  While these depend only on $\gamma$ at leading order (LO), the
contributions of the effective range $\rho=1.764$~fm to the residue,
which enter at next-to-leading order (NLO), can be expressed in terms
of the deuteron wave function renormalization constant, $Z_d$, and
treated exactly using the zed parameterization~\cite{Phillips:1999hh}.

We include the Coulomb interaction between the protons using the
t-matrix $-i t_C(E; \bm{q},\bm{p})$ for incoming (outgoing) momentum
$\bm{p}$ ($\bm{q})$. It can be expressed in closed form using the
momentum-space Coulomb wave function $\chi_{\bm p}^{(+)}({\bm q})$ as:
$t_C(E; \bm{q},\bm{p})= (E-{q}^2/{m_p}+i 0^+)\chi^{(+)}_{\bm p}({\bm
  q}) $. Coulomb amplitude $t_C$ includes non-perturbative resummation
of Coulomb photon exchanges.

The capture from $pp$ $p$-wave initial state receives contribution from
two sets of weak interactions. The first set constitutes the usual
Fermi and Gamow-Teller interactions:
\begin{align}
  \label{eq:Lweak_LO}
  \mathcal{L}^{({\rm FGT})}_W = -\frac{G_V}{\sqrt{2}}\left(l_{+}^0 N^\dagger \tau^{-} N + g_A\bm{l}_{+}\cdot N^\dagger{\bm\sigma} \tau^{-} N \right)\,,
\end{align}
where $G_V$ and $g_A$ are the vector and axial coupling constants, for which we use the latest Particle Data Group's~\cite{PhysRevD.98.030001} values of $1.1363(3)~\times~10^{-11}$~MeV$^{-2}$ and 1.2724(23), respectively. 
$l_+^\mu$ is the leptonic Dirac current, and
$\tau^{-}=(\tau_1-i \tau_2)/2$ is the isospin lowering operator.

The second set of interactions constitutes relativistic $p/m$ effects:
\begin{multline}
  \label{eq:Lweak_NLO}
  \mathcal{L}_W^{(\rm rel)} = \frac{G_V}{\sqrt{2}} \bigl[g_A l_+^0 N^\dagger {\bm \sigma} \cdot \frac{i \overleftrightarrow{\bm\nabla}}{ 2 m} \tau^{-} N \\
  +\bm{l}_+\cdot N^\dagger\bigl( \frac{i \overleftrightarrow{\bm\nabla}}{2 m} \tau^{-} 
   -\mu_V {\bm\sigma} \times \frac{\overline{\bm\nabla} }{2m}\tau^{-}
  \bigr) N
  \bigr]\,,
\end{multline}
where $\mu_V=(\mu_p-\mu_n)/2$ denotes the isovector magnetic moment,
$\overleftrightarrow{\bm\nabla}=\overleftarrow{\bm\nabla}-\overrightarrow{\bm\nabla}$ and
$\overline{\bm\nabla}=\overleftarrow{\bm\nabla}+\overrightarrow{\bm\nabla}$.

\begin{figure}[ht]
\begin{center}
\includegraphics[width=0.3\textwidth,clip=true]{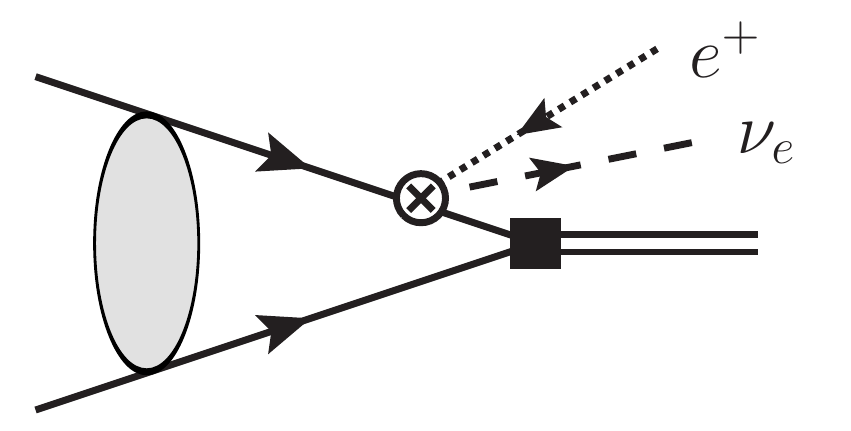} 
\end{center}
\caption{\protect Feynman diagram for $pp$ fusion: solid lines nucleons, short-dashed line positron $e^+$, dashed line neutrino $\nu_e$ and double line deuteron. The blob represents Coulomb amplitude $t_c$, ``$\otimes$'' a weak vertex,
 ``$\blacksquare$''  a strong interaction vertex.}
\label{fig:pp-weak}
\end{figure}
 
The Feynman diagrams in Fig.~\ref{fig:pp-weak} provide the dominant
$p$-wave contribution to $pp$ fusion.  A straightforward calculation
shows that $p$-wave capture from the weak interaction vertex in
Eq.~\eqref{eq:Lweak_LO} comes from the deuteron recoil momentum
$\bm{k}$. Thus this $p$-wave contribution scales as $k p/Q^2$ compared
to the LO $s$-wave amplitude in \nopi~\cite{Kong:1999tw,Kong:2000px}. We
name this recoil contribution $T_\mathrm{FGT}$.  The weak interaction
vertex generated by the $\overleftrightarrow{\bm\nabla}$ terms in
Eq.~\eqref{eq:Lweak_NLO} contribute even in the zero-recoil limit.
Relative to the LO $s$-wave amplitude, it is suppressed by a factor of
$p/m$ and we name this relativistic contribution $\Trel$.  The
contribution from the $\mu_V\overline{\bm\nabla}$ term is suppressed
by $k^2$ and we do not include it.  Compared to the LO $s$-wave
amplitude, at momentum $p\approx \gamma\approx Q$, the recoil contribution
$kp/Q^2$ and the relativistic contribution $p/m$ are similar scaling
as $Q^3/\Lambda^3\approx 0.04$. We use this estimate for
$T_\mathrm{FGT}\approx |\Trel|$ that holds up to $p\lesssim\gamma$ to
keep the EFT analysis simple. Empirically, at solar energies
$E\lesssim100$~keV, $p/m$ is small but $pk/\gamma^2$ is smaller. Thus
$\Trel$ ~contribution is larger making $p/m\lesssim0.01$ to be a
better estimate for the relative contribution of the $p$-wave amplitude.
Furthermore, the cross section (and therefore the $S$ factor) can be
decomposed into a partial wave expansion as
$\sigma(E)=\sigma_0(E)+\sigma_1(E)+\ldots$, where the subscript $l$
refers to the $l$-th $pp$ partial wave. We threfore anticipate the
$p$-wave cross section (and $S$ factor) to be smaller by a factor of
$p^2/m^2\lesssim10^{-4}$ compared to the $s$-wave value.  We include the
NLO correction from the effective range $\rho$. Initial state $p$-wave
strong interactions are suppressed by relative powers of
$Q^3/\Lambda^3$. Higher order corrections to the weak interactions are
suppressed by at least $Q^2/\Lambda^2\approx0.01$, and constitute a 10\%
uncertainty in the $p$-wave cross section.

\paragraph{\bf The $p$-wave cross section:}
\label{sec:cross_sec}
The $p$-wave amplitude is 
\begin{multline}
\label{eq:pwave_amplitude}
 i\mathcal{M}_1 = 
 i 8 \frac{G_v}{\sqrt{2}} {\epsilon_i^d}^\ast \, u_N(-\bm{p}) \mathbb{P}_j u_N(\bm{p}) \\
 \times\bigg\{\(l_+^0T_\mathrm{FGT}-\bm{l}_+\cdot\Trel\)\mathrm{Tr}\[P_i\tau^-\mathbb P_j^\dagger\]  \\
 + g_A\(l_+^kT_\mathrm{FGT}-l_+^0 T_\mathrm{rel}^k\)
  \mathrm{Tr}\[P_i\sigma_k\tau^-\mathbb P_j^\dagger\]\bigg\}\,,
\end{multline}
where $\mathbb P_j$ is the spin-triplet isospin-triplet projector,
$i\sigma_2\sigma_j(1+\tau_3)/4$. The non-relativistic two component nucleon spinor fields $u_N(\bm{p})$ are normalized as $ [u_N(\bm{p})]_\alpha[u^\ast_N(\bm{p})]_\beta=\delta_{\alpha\beta}$ when summed over polarizations. 
The amplitudes from the loop integrals are
\begin{align}
\label{eq:recoil_amp}
 T_\mathrm{FGT} = g_0 \sqrt{Z_d} m \int\frac{d^3 q}{(2\pi)^3} \chi_{\bm p}^{(+)}(\bm{q})\frac{\bm{q}\cdot\bm{k}}{\(\gamma^2+q^2\)^2}\, ,
\end{align}
and
\begin{align}
\label{eq:nlo_amp}
\Trel = g_0 \sqrt{Z_d} m \int\frac{d^3 q}{(2\pi)^3} \chi_{\bm p}^{(+)}(\bm{q}) \frac{\bm{q}}{m} \frac{1}{\gamma^2+q^2}\, .
\end{align}
The solid angle integral of $\bm{q} \chi_{\bm p}^{(+)}(\bm{q})$ picks out the vector direction $\bm{p}$ and constitutes the $l=1$ partial wave contribution. 
The c.m. deuteron momentum $\bm{k}$ is related to the positron/neutrino pair momenta $\bm{p}_{e,\nu}$ from momentum conservation as $\bm{k}=-(\bm{p}_e+\bm{p}_\nu)$. 
The 
 expressions for $T_\mathrm{FGT}\propto e^{i\delta_1}$ and
$\Trel\propto e^{i\delta_1}$ are derived further below.  Since
$T_\mathrm{FGT}\Trel^\ast=T_\mathrm{FGT}^\ast\Trel$,
Eq.~\eqref{eq:pwave_amplitude} gives
\begin{multline}
\label{eq:squared_m}
 \overline{\vert \mathcal{M}_1\vert^2} = 8 \left(\frac{G_v}{\sqrt{2}}\right)^2 
 \bigg\lbrace \(E_\nu E_e+\bm{p}_\nu\cdot\bm{p}_e\)
 \\
 \times\(3\vert T_\mathrm{FGT}\vert^2+2g_A^2\Trel\cdot\Trel^\ast\) \\
   +6\(\bm{p}_\nu\cdot\Trel\)\(\bm{p}_e\cdot\Trel^\ast\)\\
   +3\(E_\nu E_e-\bm{p}_\nu\cdot\bm{p}_e\)\Trel\cdot\Trel^\ast\\
   -\(6+4g_A^2\)\(E_\nu\bm{p}_e+E_e\bm{p}_\nu\)\cdot\Trel T_\mathrm{FGT}^\ast\\
   +2g_A^2\(3E_\nu E_e-\bm{p}_\nu\cdot\bm{p}_e\)\vert T_\mathrm{FGT}\vert^2\bigg\rbrace\,,
\end{multline}
where we used the polarization sum over the leptonic currents $l_+^\mu {l_+^\nu}^\dagger$.

The spin averaged cross section for non-relativistic fields is given by Fermi's Golden Rule as 
\begin{multline}
 \label{eq:crossec1}
 \sigma_1(E) = \int
  \frac{\mathrm{d}^3p_e}{(2\pi)^3}\frac{\mathrm{d}^3p_\nu}{(2\pi)^3} \frac{1}{4E_eE_\nu}\frac{1}{v_{rel}}
 \overline{\vert \mathcal{M}_1\vert^2} \\
   \times 2\pi\,\delta\left(\delta m +E-\frac{k^2}{2M_d}-E_e-E_\nu\right)~,
\end{multline} where $\delta m = 2m_p-M_d=m_p-m_n+\gamma^2 m/(m_p m_n)$. The integral can be reduced to 4-dimensions. The magnitude $p_\nu$ is constrained from the Dirac $\delta$-function. We are free to choose the spin quantization axis ($\hat{z}$ axis) along $\bm{p}$ direction. Azimuthal symmetry of the total lepton momentum $\bm{p}_e+\bm{p}_\nu=-\bm{k}$  implies dependence only on the difference in the azimuthal angle $\phi=\phi_e-\phi_\nu$ of the pair $\bm{p}_{e,\nu}$. 
The integral in Eq.~\eqref{eq:crossec1} can then be written as 
\begin{multline}
 \label{eq:crossec2}
 \sigma_1(E) = \frac{1}{(2\pi)^4}  \int_0^{p_e^\mathrm{max}} \mathrm{d}p_e p_e^2 \int_{-1}^{1}\mathrm{d}x_e \int_{-1}^{1}\mathrm{d}x_\nu \\
 \times\int_{0}^{2\pi} \mathrm{d}\phi 
 \frac{p_\nu^2}{\vert 1+\frac{p_\nu}{M_d}+\frac{p_e}{M_d}x_{e\nu}\vert}
 \frac{1}{4E_eE_\nu} \frac{1}{v_{rel}} \overline{\vert \mathcal{M}_1\vert^2}\,,
\end{multline}
where $x_{e,\nu}= \hat{\bm p}\cdot\hat{\bm p}_{e,\nu}$ and 
$x_{e\nu}=  \hat{\bm p}_e\cdot \hat{\bm p}_\nu=x_e x_\nu + \sqrt{1-x_e^2} \sqrt{1-x_\nu^2} \cos\phi $.
 The neutrino momentum magnitude is given by
\begin{multline}
p_\nu = -M_d - p_e x_{e\nu} + [(M_d+p_e x_{e\nu})^2\\
+2M_d(2m_p-M_d+E-E_e)
  -p_e^2]^{1/2}\,,
\end{multline}
and the maximal positron momentum is
\begin{multline}
  p_e^\mathrm{max} = \bigg\{\left(2-\frac{2m_p+E}{M_d}\right)\\
  \times \left([2m_p-M_d+E]^2 -m_e^2\right)\bigg\}^{1/2}\,.
\end{multline}

\paragraph{\bf Results:}
\label{sec:results}
The cross section $\sigma_1(E)$ in Eq.~\eqref{eq:crossec2} is evaluated by numerical integration
using analytic expressions for
$T_\mathrm{FGT}$ and $\Trel$. These can be derived from the
coordinate space wavefunction
\begin{align}
\chi_{\bm{p}}^{(+)}(\bm{r}) & \equiv
\int \frac{d^3q}{(2\pi)^3} e^{i\mathbf{q}\cdot\bm{r}} \chi_{\bm{p}}^{(+)}(\bm{q}) \nonumber\\
& =\sum_{l=0}^{\infty}(2l+1)i^{l}e^{i\delta_{l}}
P_{l}(\hat{\bm r}\cdot\hat{\bm{p}})\frac{F_{l}(\eta_p;pr)}{pr}\,,
\end{align}
where $\delta_{l}=\operatorname{arg}\Gamma(l+1+i\eta_p)$ is the Coulomb phase shift and
\begin{align}
F_l(\eta_p;\rho)=& \frac{2^l e^{-\pi\eta_p/2}|\Gamma(l+1+i\eta_p)|}
 {\Gamma(2l+2)} \,  \rho^{l+1} \, e^{-i\rho}\nonumber\\
&\times  M(l+1-i\eta_p, 2l+2,2i\rho) \, ,
\end{align}
 is the regular Coulomb wave function expressed in terms of the conventionally defined Kummer's function $M(a,b,z)$.
Equation~\eqref{eq:recoil_amp} can then be written as
\begin{align}
  T_\mathrm{FGT}=&  \frac{1}{6}g_0 \sqrt{Z_d} m ( \bm{p}\cdot\bm{k}) e^{i\delta_1} e^{-\pi\eta_p/2} |\Gamma(2+i\eta_p)|\nonumber\\
  &\times\int_0^\infty dr r^3  e^{-(\gamma+ip)r }M(2-i\eta_p,4, i  2 pr)\nonumber\\
  =&-\sqrt{\frac{8\pi\gamma}{1-\rho\gamma}} e^{i\delta_1} e^{-\pi\eta_p/2} (\bm{p}_\nu+\bm{p}_e) \cdot\bm{p}\nonumber\\
                 & \times|\Gamma(2+i\eta_p)| 
\frac{1}{(\gamma^2+p^2)^2} e^{2\eta_p \arctan{p/\gamma}}\, ,
\end{align}
where we have used the NLO relation $ g_0\sqrt{Z_d} m = \sqrt{8\pi\gamma/(1-\rho\gamma)}$. 
Similarly, Eq.~\eqref{eq:nlo_amp} can be written as
\begin{align}
\Trel=&\frac{1}{3} g_0 \sqrt{Z_d} m e^{i\delta_1} e^{-\pi\eta_p/2} |\Gamma(2+i\eta_p)|
\frac{\bm{p}}{m} \nonumber\\ 
&\times\int_0^\infty dr r (1+\gamma r) e^{-(\gamma+i p)r} M(2-i\eta_p,4, i  2 pr) \nonumber\\
=&\sqrt{\frac{8\pi\gamma}{1-\rho\gamma}} e^{i\delta_1} e^{-\pi\eta_p/2} \frac{\bm{p}}{m}|\Gamma(2+i\eta_p)| \nonumber\\
& \times \frac{1}{2p^2+2p^2\eta_p^2} \nonumber\\ & \[1+\frac{p^2+2p\eta\gamma-\gamma^2}{\gamma^2+p^2}e^{2\eta_p \arctan{p/\gamma}}\]\,.
\end{align}

\begin{figure}
  \includegraphics[width=0.49\textwidth,clip=true]{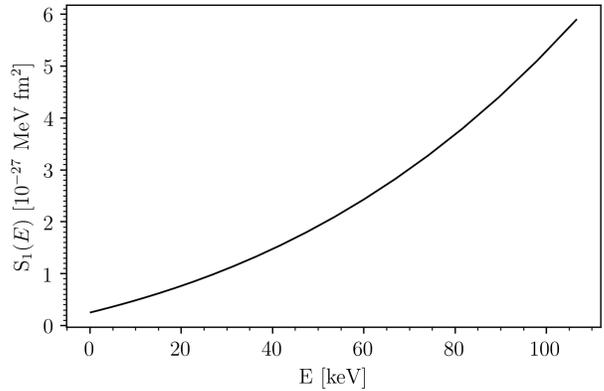}
  \caption{\label{fig:sfactor} The $p$-wave $S$-factor $S_1(E)$. }
\end{figure}

In Fig.~\ref{fig:sfactor} we show the result for the $S$-factor $S_1(E)$. We perform a polynomial fit to
the results shown in Fig.~\ref{fig:sfactor} and use it to
extrapolate the $S$ factor and its derivative to zero energy. We obtain
\begin{align}
\nonumber
  \label{eq:S1threshold}
  S_1(0) & =(2.47\pm 0.25 \pm 0.01)\times 10^{-28}\ \mathrm{ MeV~fm}^2\, ,\\
  S^\prime_1(0) & = (2.16\pm 0.22 \pm 0.01)\times 10^{-26}\ \mathrm{ fm}^2\,,
\end{align}
where the first errors indicate EFT uncertainties and the second ones
are numerical errors from polynomial fits to $S(E)$.

Our result for $S_1(0)$ agrees with the tentative estimates we made
earlier based on the power counting, but does not agree with the value
of $S_1(0)=2.0\times 10^{-25}~{\rm MeV~fm}^2$ claimed in
Reference~\cite{Marcucci:2013tda}. In fact, the $p$-wave contribution is much
smaller than the $\approx1\%$ contribution obtained by
Reference~\cite{Marcucci:2013tda} in the entire energy
region in which they perform their calculations. We, therefore,
disagree with the findings of Marcucci {\it et al.} in
Reference~\cite{Marcucci:2013tda} and claim that the $p$-wave contributions
need not be considered in the calculation of the {\it pp} $S$ factor at
astrophysically relevant energies since these are much smaller than
the precision of the $s$-wave calculation [see
Reference~\cite{Acharya:2016kfl} for a state-of-the-art uncertainty
analysis]. Furthermore, Refs.~\cite{Acharya:2016kfl,Acharya:2016rek}
have found that basis truncation errors accounted for a reduction in
Reference~\cite{Marcucci:2013tda}'s $s$-wave $S$ factor by about 0.7~\%. Since Marcucci {\it et al.} only addressed the error in the $p$-wave calculation in their Erratum~\cite{Phys.Rev.Lett.123.019901}, their revised value for the $S$ factor,
with combined $s$ and $p$ waves, is still incorrect, and does not agree with 
value
$S(0) = (4.047^{+0.024}_{-0.032}) \times 10^{-23}~{\rm MeV~fm}^2$ calulated by Reference~\cite{Acharya:2016kfl} within the uncertainty
band, which remains unmodified upon inclusion of the $p$-wave
contribution calculated in this Rapid Communication.\footnote{The relationship between the chiral EFT counterterms $c_D$ and $d_R$ have since been updated~\cite{0000000002995761}. This correction makes a negligible modification in the $S$-factor value compared to the uncertainty band~\cite{PhysRevC.98.065506}.} Finally, we emphasize that, even
though the $p$-wave numbers we calculated are negligible given the large
uncertainty in the $s$-wave value, the correction we make to
Reference~\cite{Marcucci:2013tda}'s results is at least as important as all
sources of uncertainty combined.

\paragraph{\bf Conclusion:}
\label{sec:conclusion}
We calculated for the first time the contribution of $p$-wave $pp$
configuration to the fusion rate in \nopi. This analysis was motivated
by a recent calculation with chiral potentials that suggested that the
leading $p$-wave contributions are comparable to the
next-to-next-to-leading $s$-wave contributions.

We determined the dominant Feynman diagrams contributing to the $p$-wave
$S$ factor and calculated their contribution at low energies.  The NLO
calculation includes the recoil contributions from the Gamow-Teller
and Fermi operators as well as the relativistic $1/m$ suppressed weak
interaction operators. We found that the $p$-wave contribution to the
{\it pp} fusion $S$ factor is smaller than the value obtained in 
Reference~\cite{Marcucci:2013tda} by several orders of magnitude, and that
the effect of $p$-wave fusion is therefore negligible for a high-precision determination.

Our analytic results for the $p$-wave fusion matrix element can serve as
a benchmark for numerical calculations of chiral effective theory
matrix elements. They are expressed in terms of the weak couplings
constants $g_A$ and $G_F$ and two observables from the strong sector,
the deuteron binding energy and wave function renormalization and
therefore do not suffer from any short-distance model ambiguities. We
note that the input observables needed to predict the $p$-wave fusion
rate are $s$-wave observables. Our results illustrate furthemore the
importance of calculating astrophysically relevant three-nucleon
capture reactions in pionless effective field theory to reduce
currently accepted uncertainties and to explore the importance of
recoil corrections in these nuclei.

\section{Acknowledgements}
This work was supported by the U.S. National Science Foundation under
Grants PHY-1555030 and PHY-1615092, by the Office of Nuclear Physics,
U.S. Department of Energy under Contract No. DE-AC05-00OR22725, and by
the Deutsche Forschungsgemeinschaft through The Low-Energy Frontier of the Standard Model CRC (SFB
1044) and through the PRISMA$^+$ Cluster of Excellence. GR
acknowledges support from the JINPA at Oak Ridge National Laboratory
and The University of Tennessee, during his sabbatical where part of
this research was completed.  LP and GR thank the GSI-funded EMMI RRTF
workshop ``ER15-02: Systematic Treatment of the Coulomb Interaction in
Few-Body Systems'' participants for valuable discussions and hosts for
their hospitality.



\begin{thebibliography}{18}%
\makeatletter
\providecommand \@ifxundefined [1]{%
 \@ifx{#1\undefined}
}%
\providecommand \@ifnum [1]{%
 \ifnum #1\expandafter \@firstoftwo
 \else \expandafter \@secondoftwo
 \fi
}%
\providecommand \@ifx [1]{%
 \ifx #1\expandafter \@firstoftwo
 \else \expandafter \@secondoftwo
 \fi
}%
\providecommand \natexlab [1]{#1}%
\providecommand \enquote  [1]{``#1''}%
\providecommand \bibnamefont  [1]{#1}%
\providecommand \bibfnamefont [1]{#1}%
\providecommand \citenamefont [1]{#1}%
\providecommand \href@noop [0]{\@secondoftwo}%
\providecommand \href [0]{\begingroup \@sanitize@url \@href}%
\providecommand \@href[1]{\@@startlink{#1}\@@href}%
\providecommand \@@href[1]{\endgroup#1\@@endlink}%
\providecommand \@sanitize@url [0]{\catcode `\\12\catcode `\$12\catcode
  `\&12\catcode `\#12\catcode `\^12\catcode `\_12\catcode `\%12\relax}%
\providecommand \@@startlink[1]{}%
\providecommand \@@endlink[0]{}%
\providecommand \url  [0]{\begingroup\@sanitize@url \@url }%
\providecommand \@url [1]{\endgroup\@href {#1}{\urlprefix }}%
\providecommand \urlprefix  [0]{URL }%
\providecommand \Eprint [0]{\href }%
\providecommand \doibase [0]{http://dx.doi.org/}%
\providecommand \selectlanguage [0]{\@gobble}%
\providecommand \bibinfo  [0]{\@secondoftwo}%
\providecommand \bibfield  [0]{\@secondoftwo}%
\providecommand \translation [1]{[#1]}%
\providecommand \BibitemOpen [0]{}%
\providecommand \bibitemStop [0]{}%
\providecommand \bibitemNoStop [0]{.\EOS\space}%
\providecommand \EOS [0]{\spacefactor3000\relax}%
\providecommand \BibitemShut  [1]{\csname bibitem#1\endcsname}%
\let\auto@bib@innerbib\@empty
\bibitem [{\citenamefont {Bethe}\ and\ \citenamefont
  {Critchfield}(1938)}]{Bethe:1938yy}%
  \BibitemOpen
  \bibfield  {author} {\bibinfo {author} {\bibfnamefont {H.~A.}\ \bibnamefont
  {Bethe}}\ and\ \bibinfo {author} {\bibfnamefont {C.~L.}\ \bibnamefont
  {Critchfield}},\ }\href {\doibase 10.1103/PhysRev.54.248} {\bibfield
  {journal} {\bibinfo  {journal} {Phys. Rev.}\ }\textbf {\bibinfo {volume}
  {54}},\ \bibinfo {pages} {248} (\bibinfo {year} {1938})}\BibitemShut
  {NoStop}%
\bibitem [{\citenamefont {Adelberger}\ \emph {et~al.}(1998)\citenamefont
  {Adelberger} \emph {et~al.}}]{Adelberger:1998qm}%
  \BibitemOpen
  \bibfield  {author} {\bibinfo {author} {\bibfnamefont {E.~G.}\ \bibnamefont
  {Adelberger}} \emph {et~al.},\ }\href {\doibase 10.1103/RevModPhys.70.1265}
  {\bibfield  {journal} {\bibinfo  {journal} {Rev. Mod. Phys.}\ }\textbf
  {\bibinfo {volume} {70}},\ \bibinfo {pages} {1265} (\bibinfo {year}
  {1998})},\ \Eprint {http://arxiv.org/abs/astro-ph/9805121}
  {arXiv:astro-ph/9805121 [astro-ph]} \BibitemShut {NoStop}%
\bibitem [{\citenamefont {Adelberger}\ \emph {et~al.}(2011)\citenamefont
  {Adelberger} \emph {et~al.}}]{Adelberger:2010qa}%
  \BibitemOpen
  \bibfield  {author} {\bibinfo {author} {\bibfnamefont {E.~G.}\ \bibnamefont
  {Adelberger}} \emph {et~al.},\ }\href {\doibase 10.1103/RevModPhys.83.195}
  {\bibfield  {journal} {\bibinfo  {journal} {Rev. Mod. Phys.}\ }\textbf
  {\bibinfo {volume} {83}},\ \bibinfo {pages} {195} (\bibinfo {year} {2011})},\
  \Eprint {http://arxiv.org/abs/1004.2318} {arXiv:1004.2318 [nucl-ex]}
  \BibitemShut {NoStop}%
\bibitem [{\citenamefont {Vinyoles}\ \emph {et~al.}(2017)\citenamefont
  {Vinyoles}, \citenamefont {Serenelli}, \citenamefont {Villante},
  \citenamefont {Basu}, \citenamefont {Bergström}, \citenamefont
  {Gonzalez-Garcia}, \citenamefont {Maltoni}, \citenamefont {Peña-Garay},\
  and\ \citenamefont {Song}}]{Vinyoles:2016djt}%
  \BibitemOpen
  \bibfield  {author} {\bibinfo {author} {\bibfnamefont {N.}~\bibnamefont
  {Vinyoles}}, \bibinfo {author} {\bibfnamefont {A.~M.}\ \bibnamefont
  {Serenelli}}, \bibinfo {author} {\bibfnamefont {F.~L.}\ \bibnamefont
  {Villante}}, \bibinfo {author} {\bibfnamefont {S.}~\bibnamefont {Basu}},
  \bibinfo {author} {\bibfnamefont {J.}~\bibnamefont {Bergström}}, \bibinfo
  {author} {\bibfnamefont {M.~C.}\ \bibnamefont {Gonzalez-Garcia}}, \bibinfo
  {author} {\bibfnamefont {M.}~\bibnamefont {Maltoni}}, \bibinfo {author}
  {\bibfnamefont {C.}~\bibnamefont {Peña-Garay}}, \ and\ \bibinfo {author}
  {\bibfnamefont {N.}~\bibnamefont {Song}},\ }\href {\doibase
  10.3847/1538-4357/835/2/202} {\bibfield  {journal} {\bibinfo  {journal}
  {Astrophys. J.}\ }\textbf {\bibinfo {volume} {835}},\ \bibinfo {pages} {202}
  (\bibinfo {year} {2017})},\ \Eprint {http://arxiv.org/abs/1611.09867}
  {arXiv:1611.09867 [astro-ph.SR]} \BibitemShut {NoStop}%
\bibitem [{\citenamefont {Chen}\ \emph {et~al.}(2013)\citenamefont {Chen},
  \citenamefont {Liu},\ and\ \citenamefont {Yu}}]{Chen:2012hm}%
  \BibitemOpen
  \bibfield  {author} {\bibinfo {author} {\bibfnamefont {J.-W.}\ \bibnamefont
  {Chen}}, \bibinfo {author} {\bibfnamefont {C.~P.}\ \bibnamefont {Liu}}, \
  and\ \bibinfo {author} {\bibfnamefont {S.-H.}\ \bibnamefont {Yu}},\ }\href
  {\doibase 10.1016/j.physletb.2013.02.019} {\bibfield  {journal} {\bibinfo
  {journal} {Phys. Lett.}\ }\textbf {\bibinfo {volume} {B720}},\ \bibinfo
  {pages} {385} (\bibinfo {year} {2013})},\ \Eprint
  {http://arxiv.org/abs/1209.2552} {arXiv:1209.2552 [nucl-th]} \BibitemShut
  {NoStop}%
\bibitem [{\citenamefont {Acharya}\ \emph {et~al.}(2016)\citenamefont
  {Acharya}, \citenamefont {Carlsson}, \citenamefont {Ekstr{\"o}m},
  \citenamefont {Forss{\'e}n},\ and\ \citenamefont
  {Platter}}]{Acharya:2016kfl}%
  \BibitemOpen
  \bibfield  {author} {\bibinfo {author} {\bibfnamefont {B.}~\bibnamefont
  {Acharya}}, \bibinfo {author} {\bibfnamefont {B.~D.}\ \bibnamefont
  {Carlsson}}, \bibinfo {author} {\bibfnamefont {A.}~\bibnamefont
  {Ekstr{\"o}m}}, \bibinfo {author} {\bibfnamefont {C.}~\bibnamefont
  {Forss{\'e}n}}, \ and\ \bibinfo {author} {\bibfnamefont {L.}~\bibnamefont
  {Platter}},\ }\href {\doibase 10.1016/j.physletb.2016.07.032} {\bibfield
  {journal} {\bibinfo  {journal} {Phys. Lett.}\ }\textbf {\bibinfo {volume}
  {B760}},\ \bibinfo {pages} {584} (\bibinfo {year} {2016})},\ \Eprint
  {http://arxiv.org/abs/1603.01593} {arXiv:1603.01593 [nucl-th]} \BibitemShut
  {NoStop}%
\bibitem [{\citenamefont {Marcucci}\ \emph {et~al.}(2013)\citenamefont
  {Marcucci}, \citenamefont {Schiavilla},\ and\ \citenamefont
  {Viviani}}]{Marcucci:2013tda}%
  \BibitemOpen
  \bibfield  {author} {\bibinfo {author} {\bibfnamefont {L.~E.}\ \bibnamefont
  {Marcucci}}, \bibinfo {author} {\bibfnamefont {R.}~\bibnamefont
  {Schiavilla}}, \ and\ \bibinfo {author} {\bibfnamefont {M.}~\bibnamefont
  {Viviani}},\ }\href {\doibase 10.1103/PhysRevLett.110.192503} {\bibfield
  {journal} {\bibinfo  {journal} {Phys. Rev. Lett.}\ }\textbf {\bibinfo
  {volume} {110}},\ \bibinfo {pages} {192503} (\bibinfo {year} {2013})},\
  \Eprint {http://arxiv.org/abs/1303.3124} {arXiv:1303.3124 [nucl-th]}
  \BibitemShut {NoStop}%
  \bibitem [{\citenamefont {Marcucci}\ \emph {et~al.}(2013)\citenamefont
  {Marcucci}, \citenamefont {Schiavilla},\ and\ \citenamefont
  {Viviani}}]{Phys.Rev.Lett.123.019901}%
  \BibitemOpen
  \bibfield  {author} {\bibinfo {author} {\bibfnamefont {L.~E.}\ \bibnamefont
  {Marcucci}}, \bibinfo {author} {\bibfnamefont {R.}~\bibnamefont
  {Schiavilla}}, \ and\ \bibinfo {author} {\bibfnamefont {M.}~\bibnamefont
  {Viviani}},\ }\href {\doibase 10.1103/PhysRevLett.123.019901} {\bibfield
  {journal} {\bibinfo  {journal} {Phys. Rev. Lett.}\ }\textbf {\bibinfo
  {volume} {123}},\ \bibinfo {pages} {019901(E)} (\bibinfo {year} {2019})}\
  \BibitemShut {NoStop}%
\bibitem [{\citenamefont {van Kolck}(1999)}]{vanKolck:1998bw}%
  \BibitemOpen
  \bibfield  {author} {\bibinfo {author} {\bibfnamefont {U.}~\bibnamefont {van
  Kolck}},\ }\href {\doibase 10.1016/S0375-9474(98)00612-5} {\bibfield
  {journal} {\bibinfo  {journal} {Nucl. Phys.}\ }\textbf {\bibinfo {volume}
  {A645}},\ \bibinfo {pages} {273} (\bibinfo {year} {1999})},\ \Eprint
  {http://arxiv.org/abs/nucl-th/9808007} {arXiv:nucl-th/9808007} \BibitemShut
  {NoStop}%
\bibitem [{\citenamefont {Chen}\ \emph {et~al.}(1999)\citenamefont {Chen},
  \citenamefont {Rupak},\ and\ \citenamefont {Savage}}]{Chen_1999}%
  \BibitemOpen
  \bibfield  {author} {\bibinfo {author} {\bibfnamefont {J.-W.}\ \bibnamefont
  {Chen}}, \bibinfo {author} {\bibfnamefont {G.}~\bibnamefont {Rupak}}, \ and\
  \bibinfo {author} {\bibfnamefont {M.~J.}\ \bibnamefont {Savage}},\ }\href
  {\doibase 10.1016/s0375-9474(99)00298-5} {\bibfield  {journal} {\bibinfo
  {journal} {Nuclear Physics A}\ }\textbf {\bibinfo {volume} {653}},\ \bibinfo
  {pages} {386–412} (\bibinfo {year} {1999})}\BibitemShut {NoStop}%
\bibitem [{\citenamefont {Kong}\ and\ \citenamefont
  {Ravndal}(1999)}]{Kong:1999tw}%
  \BibitemOpen
  \bibfield  {author} {\bibinfo {author} {\bibfnamefont {X.}~\bibnamefont
  {Kong}}\ and\ \bibinfo {author} {\bibfnamefont {F.}~\bibnamefont {Ravndal}},\
  }\href {\doibase 10.1016/S0375-9474(99)00314-0} {\bibfield  {journal}
  {\bibinfo  {journal} {Nucl. Phys.}\ }\textbf {\bibinfo {volume} {A656}},\
  \bibinfo {pages} {421} (\bibinfo {year} {1999})},\ \Eprint
  {http://arxiv.org/abs/nucl-th/9902064} {arXiv:nucl-th/9902064 [nucl-th]}
  \BibitemShut {NoStop}%
\bibitem [{\citenamefont {Butler}\ and\ \citenamefont
  {Chen}(2001)}]{Butler:2001jj}%
  \BibitemOpen
  \bibfield  {author} {\bibinfo {author} {\bibfnamefont {M.}~\bibnamefont
  {Butler}}\ and\ \bibinfo {author} {\bibfnamefont {J.-W.}\ \bibnamefont
  {Chen}},\ }\href {\doibase 10.1016/S0370-2693(01)01152-2} {\bibfield
  {journal} {\bibinfo  {journal} {Phys. Lett.}\ }\textbf {\bibinfo {volume}
  {B520}},\ \bibinfo {pages} {87} (\bibinfo {year} {2001})},\ \Eprint
  {http://arxiv.org/abs/nucl-th/0101017} {arXiv:nucl-th/0101017 [nucl-th]}
  \BibitemShut {NoStop}%
\bibitem [{\citenamefont {Kaplan}\ \emph
  {et~al.}(1998{\natexlab{a}})\citenamefont {Kaplan}, \citenamefont {Savage},\
  and\ \citenamefont {Wise}}]{Kaplan:1998we}%
  \BibitemOpen
  \bibfield  {author} {\bibinfo {author} {\bibfnamefont {D.~B.}\ \bibnamefont
  {Kaplan}}, \bibinfo {author} {\bibfnamefont {M.~J.}\ \bibnamefont {Savage}},
  \ and\ \bibinfo {author} {\bibfnamefont {M.~B.}\ \bibnamefont {Wise}},\
  }\href@noop {} {\bibfield  {journal} {\bibinfo  {journal} {Nucl. Phys. B}\
  }\textbf {\bibinfo {volume} {534}},\ \bibinfo {pages} {329} (\bibinfo {year}
  {1998}{\natexlab{a}})},\ \Eprint {http://arxiv.org/abs/nucl-th/9802075}
  {nucl-th/9802075} \BibitemShut {NoStop}%
\bibitem [{\citenamefont {Kaplan}\ \emph
  {et~al.}(1998{\natexlab{b}})\citenamefont {Kaplan}, \citenamefont {Savage},\
  and\ \citenamefont {Wise}}]{Kaplan:1998tg}%
  \BibitemOpen
  \bibfield  {author} {\bibinfo {author} {\bibfnamefont {D.~B.}\ \bibnamefont
  {Kaplan}}, \bibinfo {author} {\bibfnamefont {M.~J.}\ \bibnamefont {Savage}},
  \ and\ \bibinfo {author} {\bibfnamefont {M.~B.}\ \bibnamefont {Wise}},\
  }\href {\doibase 10.1016/S0370-2693(98)00210-X} {\bibfield  {journal}
  {\bibinfo  {journal} {Phys. Lett.}\ }\textbf {\bibinfo {volume} {B424}},\
  \bibinfo {pages} {390} (\bibinfo {year} {1998}{\natexlab{b}})},\ \Eprint
  {http://arxiv.org/abs/nucl-th/9801034} {arXiv:nucl-th/9801034} \BibitemShut
  {NoStop}%
\bibitem [{\citenamefont {Kaplan}\ \emph {et~al.}(1999)\citenamefont {Kaplan},
  \citenamefont {Savage},\ and\ \citenamefont {Wise}}]{KSW99}%
  \BibitemOpen
  \bibfield  {author} {\bibinfo {author} {\bibfnamefont {D.~B.}\ \bibnamefont
  {Kaplan}}, \bibinfo {author} {\bibfnamefont {M.~J.}\ \bibnamefont {Savage}},
  \ and\ \bibinfo {author} {\bibfnamefont {M.~B.}\ \bibnamefont {Wise}},\
  }\href {\doibase 10.1103/PhysRevC.59.617} {\bibfield  {journal} {\bibinfo
  {journal} {Phys. Rev.}\ }\textbf {\bibinfo {volume} {C59}},\ \bibinfo {pages}
  {617} (\bibinfo {year} {1999})},\ \Eprint
  {http://arxiv.org/abs/nucl-th/9804032} {arXiv:nucl-th/9804032} \BibitemShut
  {NoStop}%
\bibitem [{\citenamefont {Phillips}\ \emph {et~al.}(2000)\citenamefont
  {Phillips}, \citenamefont {Rupak},\ and\ \citenamefont
  {Savage}}]{Phillips:1999hh}%
  \BibitemOpen
  \bibfield  {author} {\bibinfo {author} {\bibfnamefont {D.~R.}\ \bibnamefont
  {Phillips}}, \bibinfo {author} {\bibfnamefont {G.}~\bibnamefont {Rupak}}, \
  and\ \bibinfo {author} {\bibfnamefont {M.~J.}\ \bibnamefont {Savage}},\
  }\href {\doibase 10.1016/S0370-2693(99)01496-3} {\bibfield  {journal}
  {\bibinfo  {journal} {Phys. Lett.}\ }\textbf {\bibinfo {volume} {B473}},\
  \bibinfo {pages} {209} (\bibinfo {year} {2000})},\ \Eprint
  {http://arxiv.org/abs/nucl-th/9908054} {arXiv:nucl-th/9908054} \BibitemShut
  {NoStop}%
\bibitem [{\citenamefont {Tanabashi}\ \emph {et~al.}(2018)\citenamefont
  {Tanabashi} \emph {et~al.}}]{PhysRevD.98.030001}%
  \BibitemOpen
  \bibfield  {author} {\bibinfo {author} {\bibfnamefont {M.}~\bibnamefont
  {Tanabashi}} \emph {et~al.} (\bibinfo {collaboration} {Particle Data
  Group}),\ }\href {\doibase 10.1103/PhysRevD.98.030001} {\bibfield  {journal}
  {\bibinfo  {journal} {Phys. Rev. D}\ }\textbf {\bibinfo {volume} {98}},\
  \bibinfo {pages} {030001} (\bibinfo {year} {2018})}\BibitemShut {NoStop}%
\bibitem [{\citenamefont {Kong}\ and\ \citenamefont
  {Ravndal}(2001)}]{Kong:2000px}%
  \BibitemOpen
  \bibfield  {author} {\bibinfo {author} {\bibfnamefont {X.}~\bibnamefont
  {Kong}}\ and\ \bibinfo {author} {\bibfnamefont {F.}~\bibnamefont {Ravndal}},\
  }\href {\doibase 10.1103/PhysRevC.64.044002} {\bibfield  {journal} {\bibinfo
  {journal} {Phys. Rev.}\ }\textbf {\bibinfo {volume} {C64}},\ \bibinfo {pages}
  {044002} (\bibinfo {year} {2001})},\ \Eprint
  {http://arxiv.org/abs/nucl-th/0004038} {arXiv:nucl-th/0004038 [nucl-th]}
  \BibitemShut {NoStop}%
\bibitem [{\citenamefont {Acharya}\ \emph {et~al.}(2017)\citenamefont
  {Acharya}, \citenamefont {Ekstr{\"o}m}, \citenamefont {Odell}, \citenamefont
  {Papenbrock},\ and\ \citenamefont {Platter}}]{Acharya:2016rek}%
  \BibitemOpen
  \bibfield  {author} {\bibinfo {author} {\bibfnamefont {B.}~\bibnamefont
  {Acharya}}, \bibinfo {author} {\bibfnamefont {A.}~\bibnamefont
  {Ekstr{\"o}m}}, \bibinfo {author} {\bibfnamefont {D.}~\bibnamefont {Odell}},
  \bibinfo {author} {\bibfnamefont {T.}~\bibnamefont {Papenbrock}}, \ and\
  \bibinfo {author} {\bibfnamefont {L.}~\bibnamefont {Platter}},\ }\href
  {\doibase 10.1103/PhysRevC.95.031301} {\bibfield  {journal} {\bibinfo
  {journal} {Phys. Rev.}\ }\textbf {\bibinfo {volume} {C95}},\ \bibinfo {pages}
  {031301(R)} (\bibinfo {year} {2017})},\ \Eprint
  {http://arxiv.org/abs/1608.04699} {arXiv:1608.04699 [nucl-th]} \BibitemShut
  {NoStop}%
\bibitem [{\citenamefont {Gazit}\ \emph {et~al.}(2008)\citenamefont {Gazit},
  \citenamefont {Quaglioni},\ and\ \citenamefont
  {Navratil}}]{0000000002995761}%
  \BibitemOpen
  \bibfield  {author} {\bibinfo {author} {\bibfnamefont {D.}~\bibnamefont
  {Gazit}}, \bibinfo {author} {\bibfnamefont {S.}~\bibnamefont {Quaglioni}}, \
  and\ \bibinfo {author} {\bibfnamefont {P.}~\bibnamefont {Navratil}},\ }\href
  {https://doi.org/10.1103/PhysRevLett.103.102502}
  {\bibfield  {journal} {\bibinfo  {journal} {Phys. Rev. Lett.}\
  }\textbf {\bibinfo {volume} {103}},\ \bibinfo {pages}{102502}  (\bibinfo {year} {2009})};~\href{https://doi.org/10.1103/PhysRevLett.122.029901}{\bibfield  {journal} \textbf {\bibinfo {volume} {122}},\ \bibinfo {pages}{029901(E)}  (\bibinfo {year} {2019}) }\BibitemShut
  {NoStop}%
\bibitem [{\citenamefont {Acharya}\ \emph {et~al.}(2018)\citenamefont
  {Acharya}, \citenamefont {Ekstr{\"o}m},\ and\ \citenamefont {Platter}}]{PhysRevC.98.065506}%
  \BibitemOpen
  \bibfield  {author} {\bibinfo {author} {\bibfnamefont {B.}~\bibnamefont
  {Acharya}}, \bibinfo {author} {\bibfnamefont {A.}~\bibnamefont
  {Ekstr{\"o}m}},
  \ and\
  \bibinfo {author} {\bibfnamefont {L.}~\bibnamefont {Platter}},\ }\href
  {\doibase 10.1103/PhysRevC.98.065506} {\bibfield  {journal} {\bibinfo
  {journal} {Phys. Rev.}\ }\textbf {\bibinfo {volume} {C98}},\ \bibinfo {pages}
  {065506} (\bibinfo {year} {2018})},\ \Eprint
  {http://arxiv.org/abs/1806.09481v2} {arXiv:1806.09481 [nucl-th].} \BibitemShut
  {NoStop}%
  
    
\end{thebibliography}
\end{document}